\documentclass[useAMS,usenatbib]{mn2e}

\usepackage{graphicx}

\title[AGN-Starburst connection in NGC\,7582]{AGN-Starburst connection in NGC\,7582: GNIRS IFU observations}

\author[Rogemar A. Riffel et al.]{Rogemar
A. Riffel$^{1}$\thanks{E-mail:rogemar@ufrgs.br}, Thaisa
Storchi-Bergmann$^{1}$, Oli L. Dors Jr$^{1}$ and \newauthor Cl\'audia
Winge$^{2}$\\
$^{1}$Universidade Federal do Rio Grande do Sul, IF, CP 15051, Porto Alegre 91501-970, RS, Brazil.\\
$^{2}$Gemini Observatory, c/o AURA Inc., Casilla 603, La Serena, Chile.}

\begin{document}

\date{Accepted 1988 December 15. Received 1988 December 14; in original form 1988 October 11}

\pagerange{\pageref{firstpage}--\pageref{lastpage}} \pubyear{2002}

\maketitle

\label{firstpage}

\begin{abstract}
We analyse two-dimensional near-IR $K-$band spectra from the inner
660$\times$315\,pc$^2$ of the Seyfert galaxy NGC\,7582 obtained with
the Gemini GNIRS IFU at a spatial resolution of $\approx$\,50\,pc and spectral resolving power R\,$\approx$5900. The nucleus harbors an unresolved source well reproduced by a blackbody of temperature T\,$\approx$1050\,K, which we attribute to emission by circumnuclear dust located closer than 25\,pc from the nucleus, with total mass of $\approx3\times\,10^{-3}$\,M$_\odot$. Surrounding the nucleus, we observe a ring of active star formation, apparently in the galactic plane, with radius of $\approx$\,190\,pc, an age of $\approx$\,5\,Myr and total mass of ionized gas of $\approx3\times\,10^6$\,M$_\odot$. The radiation of the young stars in the ring accounts for at least 80\% of the ionization observed in the Br$\gamma$ emitting gas, the remaining being due to radiation emitted by the active nucleus. 
The stellar kinematics was derived using the CO absorption band at $2.29\,\mu$m and reveals: (1) a distorted rotation pattern in the radial velocity field with kinematic center apparently displaced from the nuclear source by a few tens of parsecs; (2) a high velocity dispersion in the bulge of $\sigma_*=170$\,km\,s$^{-1}$; (3) a partial ring of $\sigma_*=50$\,km\,s$^{-1}$, located close to the Br$\gamma$ emitting ring, but displaced by $\approx$\,50\,pc towards the nucleus, interpreted as due to stars formed from cold gas in a previous burst of star formation. The kinematics of the ionized gas shows a similar rotation pattern to that of the stars, plus a blueshifted component with velocities $\ge$\,100\,km\,s$^{-1}$ interpreted as due to an outflow along the ionization cone, which was partially covered by our observations. The mass outflow rate in the ionized gas was estimated as $\dot{M}_{\rm HII}\approx0.05\,{\rm M_\odot\,yr^{-1}}$, which is one order of magnitude larger than the accretion rate to the active galactic nucleus (AGN), indicating that the outflowing gas does not originate in the AGN, but is instead the circumnuclear gas from the host galaxy being pushed away by a nuclear outflow.
The flux distribution and kinematics of the hot molecular gas, traced by the H$_2\,\lambda$\,2.22$\mu$m emission line, suggests that most of this gas is in the galactic plane. An excess blueshift along  PA$\approx-70^\circ$, where a nuclear bar has been observed, can be interpreted as an inflow towards the nucleus. We thus conclude that the H$_2$ kinematics traces the feeding of the AGN, while the ionized  gas kinematics traces its feedback via the outflows. 
An AGN-Starburst connection in the nucleus of NGC\,7582 is supported by the ratio between the mass accretion rate and the star formation rate in the circumnuclear region of $\approx$0.26\%, which is close to the expected relation between the mass of the SMBH and that of the host galaxy bulge in galaxies (the Magorrian relation).


\end{abstract}

\begin{keywords}
galaxies: Seyfert -- infrared: galaxies -- galaxies: NGC\,7582 (individual) -- galaxies: kinematics -- galaxies: outflows -- galaxies: starburst
\end{keywords}

\section{Introduction}
\label{introd.}

\begin{figure*}
\centering
\includegraphics[scale=1]{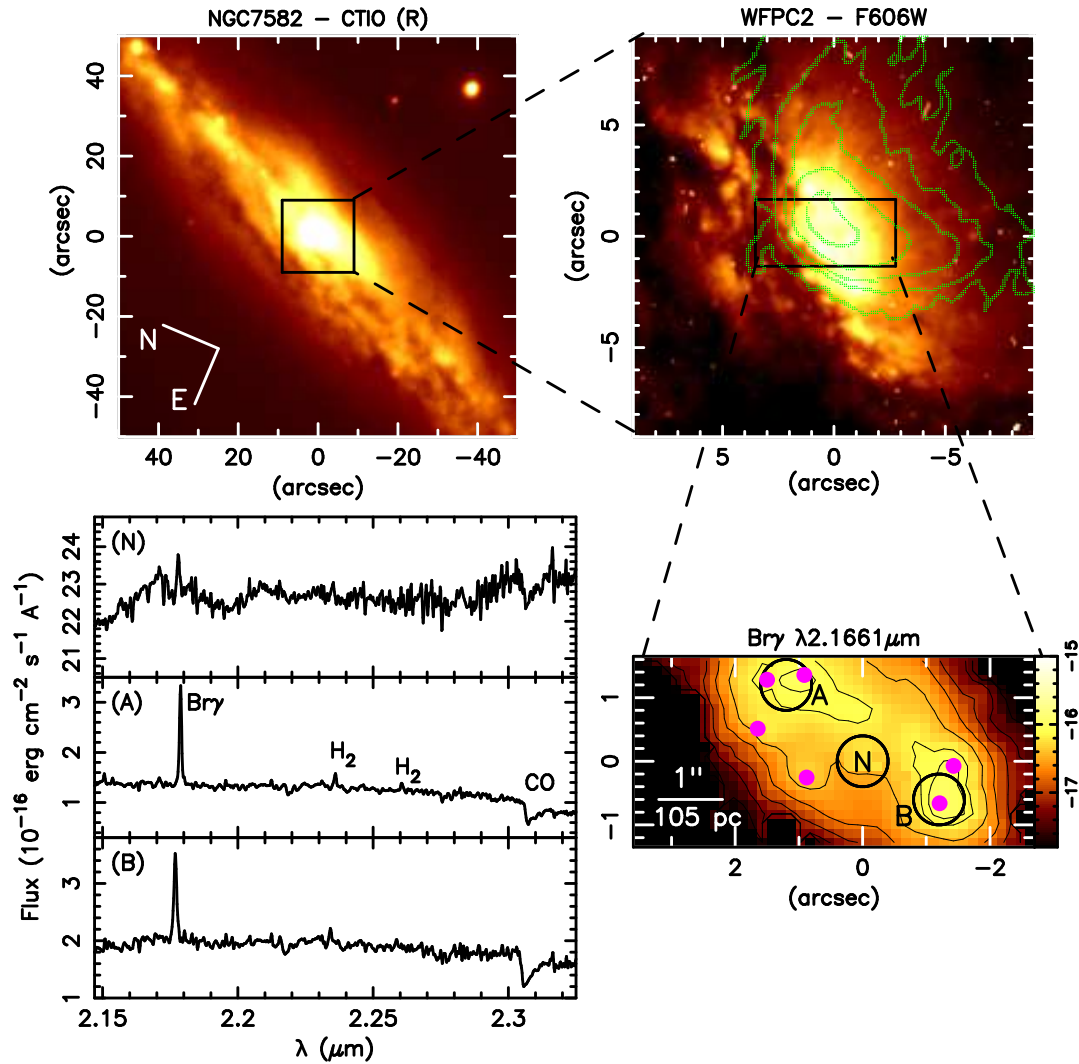}
\caption{Top left:  Large scale image of NGC\,7582 from
\citet{hameed99}. Top right: HST WFPC2 image for the central region of NGC\,7582. The green contours are for the [O\,{\sc iii}] emission form \citet{sb99}. Bottom right: Br$\gamma$ flux map reconstructed from the GNIRS IFU spectroscopy. Bottom left: 
Spectra  extracted for a circular aperture of 0$\farcs$4 radius centred at the positions N, A and B  marked at the bottom-right panel with the emission lines and CO absorption bandhead
identified.}
\label{large}
\end{figure*}

Many recent studies on the stellar population in the vicinity of active galactic nuclei (hereafter AGN) report an excess of young to intermediate age stars in the inner few hundred parsecs  when compared to non-active galaxies \citep[e.g.][]{heckman97,gonzalez-delgado98,schmitt00,cid01,sb01,kauffmann03,cid05,davies07}. This excess supports the existence of the so-called AGN-Starburst connection \citep{perry-dyson85,terlevich-melnick85,norman88, cid95}, which can be understood as due to the fact that both circumnuclear star formation and nuclear activity feed on gas inflowing towards the nuclei of galaxies. The gas infllow, if massive enough, in its way to the centre,  may trigger circumnuclear  star formation, and, reaching the nucleus, also the nuclear activity. Alternatively, there may be a delay in the triggering of the nuclear activity, which would be fed via mass loss from the evolving stars.
The AGN-Starburst connection is consistent and somewhat expected within the presently favored scenario of co-evolution of galaxies and their nuclear supermassive black holes (SMBH), supported by the M$-\sigma$ relation \citep{tremaine02,fm00,gb00}: if the bulge grows via the formation of new stars (starbursts), and the SMBH grows via mass accretion from its surroundings, one could expect a relation between the nuclear mass accretion rate and the star-formation rate in circumnuclear starbursts.

The mechanism through which the gas makes its way from the circumnuclear regions of  star formation -- which in many cases have a ring-like morphology -- to the AGN are nevertheless still elusive. In \citet{simoes-lopes07}, using optical images obtained with the Hubble Space Telescope (HST) of the inner few hundred parsecs of both active and non-active galaxies, the authors concluded that there is a strong correlation between the presence of dusty structures (spirals and filaments) and nuclear activity in early-type galaxies, proposing that these structures trace the material in its way in to feed the active nucleus. Recently, this was shown to be the case for at least two low-activity AGN with circumnuclear rings of star-formation: NGC\,1097 \citep{fathi06} and NGC\,6951 \citep{sb07}. In these two galaxies, the authors have observed streaming motions along nuclear spirals in the H$\alpha$ emitting gas. Another similar case is the Seyfert galaxy NGC\,4051 where \citet{riffel08} found streaming motions along a nuclear spiral  in molecular gas emitting the H$_2\lambda$2.1218$\mu$m line.

Inflows towards AGN are difficult to observe, particularly in ionized gas, due the predominance of emission from outflowing gas in the vicinity of the active nucleus.  The inflow probably  occurs mostly  in ``cold'' gas, while the outflows are seen in ionized gas \citep[e.g.][]{crenshaw00,das05,das06,riffel06}. Studies aimed to investigate the kinematics and flux distributions of the molecular and ionized gas in the near-infrared (hereafter near-IR) bands indeed reveal that the H$_2$ emitting gas usually has a distinct kinematics and distribution from that of the ionized gas. The H$_2$ kinematics seems to be ``colder''  and, together with the flux distribution, suggest that the molecular gas is more restricted to the galactic plane, while the ionized gas seems to extend to higher galactic latitudes and has a ``hotter'' kinematics \citep{sb99,rodriguez-ardila04,rodriguez-ardila05,riffel06,riffel08}. These studies suggest that the feeding of AGN  is dominated by inflow of cold molecular gas, while the ionized gas constributes mostly to its feedback via the outflows.

This work is aimed at investigating the relation between circumnuclear star formation and nuclear activity, as well as to look for the mechanisms of feeding and feedback in the nearby Seyfert galaxy NGC\,7582, which has a Hubble type SBab and has a distance of 21.6\,Mpc  (for $z=0.00525$ from \citet{devaucouleurs91}, with 1$\farcs$0 corresponding to $\approx$\ 105\,pc at the galaxy, adopting $H_0=73{\rm \,km\,s^{-1}\,Mpc^{-1}}$). This galaxy is an ideal candidate for this study because it has both circumnuclear star formation \citep{regan99,sb01,sosa-brito01,wold06a} and a nuclear outflow observed by \cite{morris85} using the TAURUS Fabry-Perot spectrometer. The outflows, observed in  the [O\,{\sc iii}] emission line are oriented approximately perpendicular to the galactic plane, with velocities of up to $-100\,{\rm km\,s^{-1}}$, and are co-spatial with an ionization cone observed in an  [O\,{\sc iii}]$\,\lambda5007$ narrow band image \citep{sb91}. 

Besides mapping the  [O\,{\sc iii}] outflow, \citet{morris85} also reported the presence of a kpc-scale rotating disc in the plane of the galaxy, observed in H$\alpha$. \citet{wold06a} have studied the star formation in the
circumnuclear  disc using [Ne\,{\sc ii}]12.8$\mu$m narrow band images,
and report the presence of embedded star clusters claiming that there are no counterparts detected at optical or near-IR wavelengths. They calculate the mass of the  SMBH as 5.5$\times$10$^7$\,M$_\odot$ using high resolution mid infrared spectroscopy. 

We use K-band spectroscopic observations obtained with the Integral Field Unit (IFU) of the Gemini Near-Infrared Spectrograph (GNIRS) to map the stellar and gaseous flux distributions and kinematics of the inner  few hundred parsecs of NGC\,7582. The K-band spectra allow the mapping of the stellar kinematics via the CO\,$\lambda$2.3$\mu$m absorption feature, the ionized gas flux distribution and kinematics via the Br$\gamma$ emission line and the molecular gas flux distribution and kinematics via the H$_2$ emission lines. This paper is organized as follows: in Section 2 we describe the observations and data reduction. In Section 3 we present the two-dimensional flux distributions and  kinematics for both the emitting gas and stars. The results are discussed in Section 4, and in Section 5 we present our conclusions.

\section{Observations and Data Reduction}

The data were obtained at the Gemini South Telescope with the GNIRS IFU
\citep{elias98,allington06,allington07} in 2005 Nov 10 as part of
program GS-2005B-Q-65, and comprise four individual exposures of
900\,s  centred at  $\lambda$=2.24$\mu$m. The 111 l/mm
grating with the Short Blue Camera (0$\farcs$15/pixel) was used,
resulting in a nominal resolving power of R=5900. The GNIRS IFU has a
rectangular field of view, of approximately
3$\farcs$2\,$\times$\,4$\farcs$8, divided into 21 slices. At the
detector, the slices are divided along their length into 0$\farcs$15
square IFU elements.  The major axis of the IFU was oriented along 
position angle  PA=203$^\circ$, in order to partially cover the ionization cone
observed in the [O\,{\sc iii}]$\lambda5007$ narrow-band image by
\citet{sb91}. Two sets of observations were obtained, one centred at
 1$\farcs$2 from the nucleus along PA=23$^\circ$ and another
centred at  0$\farcs$75 from the nucleus along
PA=203$^\circ$, the small offset being required to fill in the gaps
between the slices. The observing procedure followed the standard
Object--Sky--Sky--Object dither sequence, with off-source sky
positions since the target is extended. 
Conditions during the observations were stable and clear, with image
quality in $K$-band, in the range 0$\farcs$45-0$\farcs$55, as obtained for
the full width at half maximuum (FWHM) of the stellar profiles. The resulting spectral resolution, measured from the FWHM or the emission lines of the arc calibrations is  $\approx$2.9\,\AA.

The data reduction was accomplished using tasks in the {\sc
gemini.gnirs iraf} package as well as generic {\sc iraf} tasks. The
reduction procedure included trimming of the images, flat-fielding,
sky subtraction, wavelength calibration and s-distortion correction. The
telluric bands were removed and the frames flux calibrated by
interpolating a black body function to the spectrum of the telluric
standard star, observed immediatelly after the galaxy. The final IFU
datacube contains 840 spectra covering the spectral
range from 2.14$\,\mu$m to 2.33$\,\mu$m, each spectrum corresponding
to an angular coverage of 0$\farcs$15$\times$0$\farcs$15, which
translates to 16$\times$16\,pc$^2$ at the galaxy. The total observed field of
view 6$\farcs$3$\times$3$\farcs$0 (obtained by mosaicing the two set
of observations) corresponds to a region of projected dimensions
660\,pc$\times$315\,pc at the galaxy.  

\section{Results}
\label{results}

In the top-left panel of Figure\,\ref{large} we present a large scale
optical $R$-band image of NGC\,7582 from \citet{hameed99}, obtained with the Cerro Tololo 
Inter-American Observatory (CTIO) 1.5\,m telescope. The top-right panel presents 
a continuum image observed with the HST Wide Field Planetary Camera 2 (WFPC2) using the filter F606W 
obtained from the HST archive (program 8597 -- PI: Regan, M.). We have overlaid on the HST image the contours (in green) of the narrow-band [O\,{\sc iii}]\,$\lambda5007$ image from \citet{sb91}, which maps the ionization cone. The central rectangle shows the field-of-view of 
the GNIRS IFU observations, which covers only a small fraction of the ionization cone observed in [O\,{\sc iii}]. In the bottom-right panel 
we present a Br$\gamma$ flux distribution image reconstructed from the
GNIRS IFU datacube, where we have marked, in magenta, the star clusters found by \citet{wold06a}. These clusters delineate an elongated ring of recent star formation, with semi-major axis of $\approx1\farcs7$, which was almost entirely  covered by our observations.

In the bottom-left panel, we show three charateristic IFU spectra obtained within circular apertures of $0\farcs4$ radius: the nuclear spectrum -- centred on the location corresponding to the peak flux in the continuum (position N in the Br$\gamma$ intensity
map), and spectra from two locations at 1\farcs7 NW (position A), and at 1\farcs7
SE of the nucleus  (position B), located in the star-forming ring. The  spectra of the latter regions present emission lines of
Br$\gamma\,\lambda2.1661\,\mu$m, H$_2\,\lambda2.2235\,\mu$m and
H$_2\,\lambda2.2477\,\mu$m, used to map the gaseous kinematics and
flux distributions, and the stellar absorption band of
$^{12}$CO\,(2,0)\,$\lambda2.2935\,\mu$m, used to obtain the stellar
kinematics. The nuclear spectrum is redder than those from the ring, with a flux in the continuum 10$-$15 times higher than in the ring, and shows a narrow Br\,$\gamma$ emission line, on the top of what seems to be a broad component of the line.

\subsection{Emission-line flux distributions}

\begin{figure}
\includegraphics[scale=1.1]{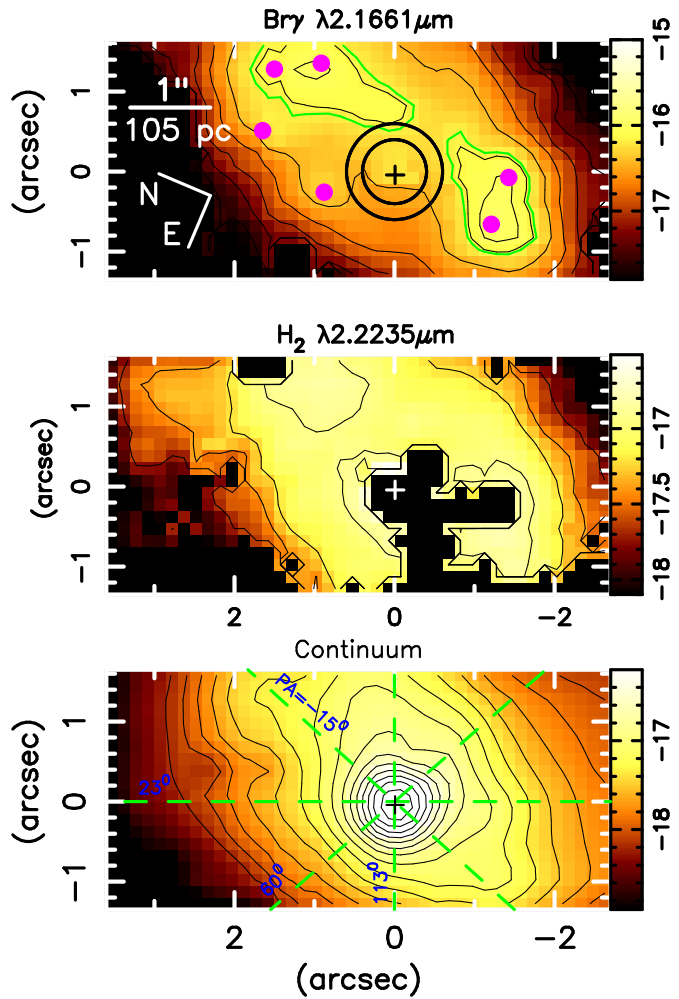}
\caption{2D flux maps for the Br$\gamma$ (top), and
H$_2\,\lambda2.2235\,\mu$m (middle) emission lines with mean
uncertainties of 6\% and 11\%, respectively, and reconstructed image for the
$2.17\,\mu$m continuum emission (bottom). The black contours
are iso-intensity curves, and the green contours in the Br$\gamma$
flux map delimit  regions where the line flux per pixel is higher than 50\% of
the peak flux in the region. The magenta filled circles indicate the position of the 
embedded star clusters observed by \citet{wold06a}. The black circles surrounding the nucleus in the Br$\gamma$ flux map represent the apertures used to extract the spectra shown in Fig.\,\ref{fitting} and the green dashed lines show the orientation of the pseudo-slits used to construct one-dimensional cuts shown in Fig.\,\ref{cut}. }

\label{flux}
\end{figure}


Flux distributions in the emission lines were obtained directly from the datacube
by integrating the flux under each  emission-line profile and subtracting the continuum contribution using images obtained in windows adjacent to the emission line. 

We have also adjusted Gaussian profiles to the emission lines
in order to measure the gas radial velocity (from the central
wavelength of the line) and velocity dispersion (from the width of the
line). The fitting of each emission-line profile was obtained using the {\sc splot iraf} 
task, and resulted also in an integrated flux for each line, from which we also obtained  flux distributions. A comparison of the flux distributions obtained via the fitting of Gaussians to the flux distributions obtained directly from the datacube showed that, although both flux distributions agree with each other within the uncertainties, the ones obtained via Gaussian fitting are less noisy away from the nucleus. These flux distributions are shown in 
Figure\,\ref{flux} for the Br$\gamma$ (top  panel) and H$_2\,\lambda2.2235\,\mu$m (middle panel) emission lines with mean uncertainties of 6\% and 11\%, respectively. 
The Br$\gamma$ emission is extended along the major
axis of the galaxy (PA$\approx165\degr$) and show flux enhancements in an arc shaped structure which approximately trace the star clusters observed by \citet{wold06a}, shown again by the magenta circles. Two strong Br$\gamma$ emission peaks in the ring are located  at $\approx1\farcs7$\,NW and $\approx1\farcs7$\,SE of the \,nucleus, and are encircled by the green contours in Fig.\,\ref{flux} which correspond to the limits within which
the flux in each pixel is higher than 50\% of the peak flux in the region. We note that each of our encircled regions include two of the clusters identified by \citet{wold06a}. While the Br$\gamma$ flux distribution show intensity variations and enhacements in the position corresponding to the star clusters, mapping the ring of star-formation, the H$_2$ flux distribution (middle panel of Fig.\,\ref{flux}) is more uniform througouth the nuclear region, showing no enhancements at the star clusters. The black regions correspond to pixels for which  the S/N ratio was too low to allow a reliable measurement of the emission line.

The bottom panel of Figure\,\ref{flux} shows the
$2.17\,\mu$m continuum image obtained from interpolation between two 
continuum windows adjacent to the Br$\gamma$ profile. The  iso-intensity contours in the continuum image are assymetric, being more extended to the W--SW than to the E--NE, as expected, due to the strong dust absorption observed in optical images, in good agreement with the 1.6\,$\mu$m continuum image presented by \citet{wold06a}.


\subsection{Stellar kinematics}\label{res-stel}

\begin{figure}
\includegraphics[scale=0.42]{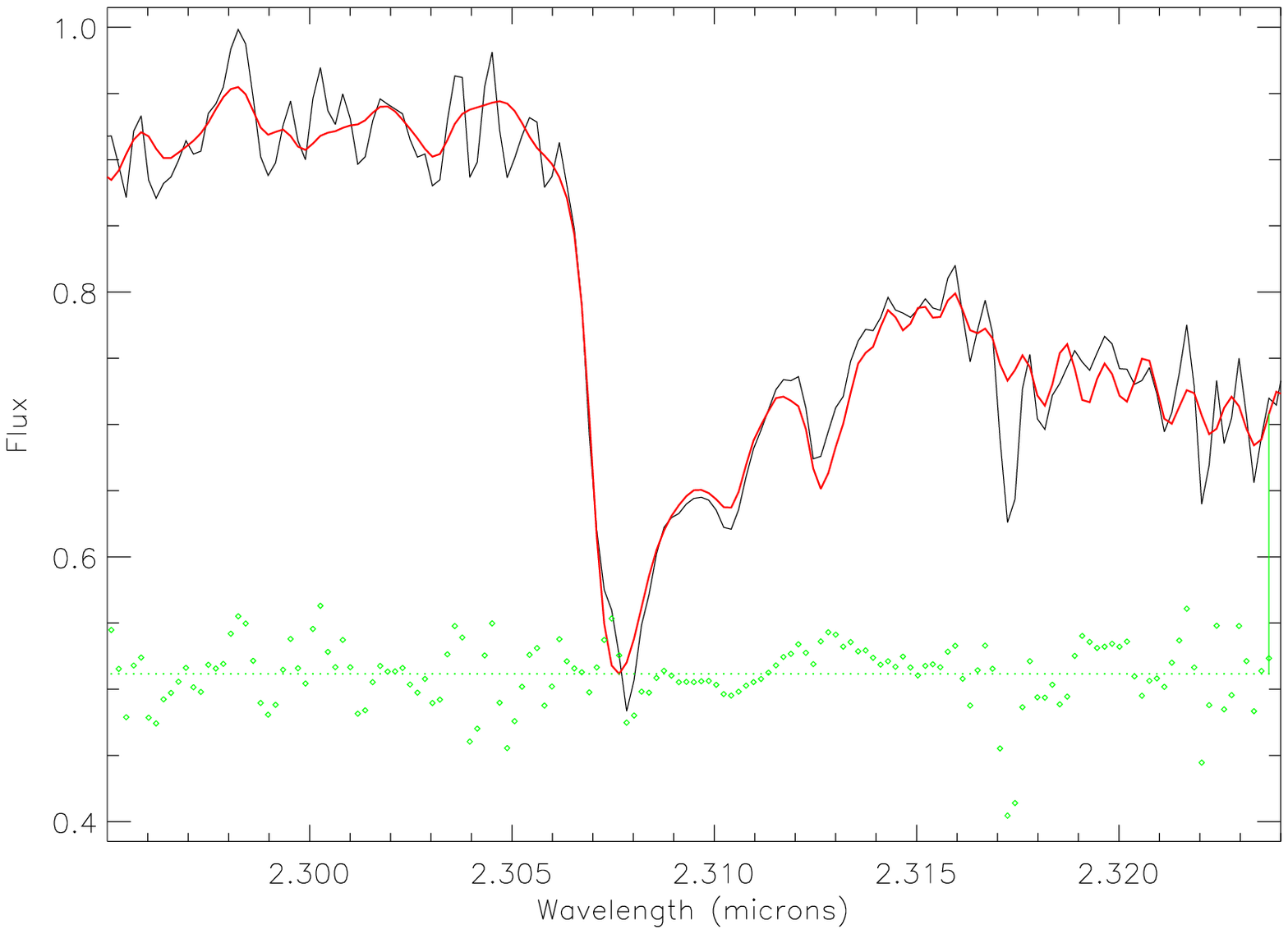}
\includegraphics[scale=0.42]{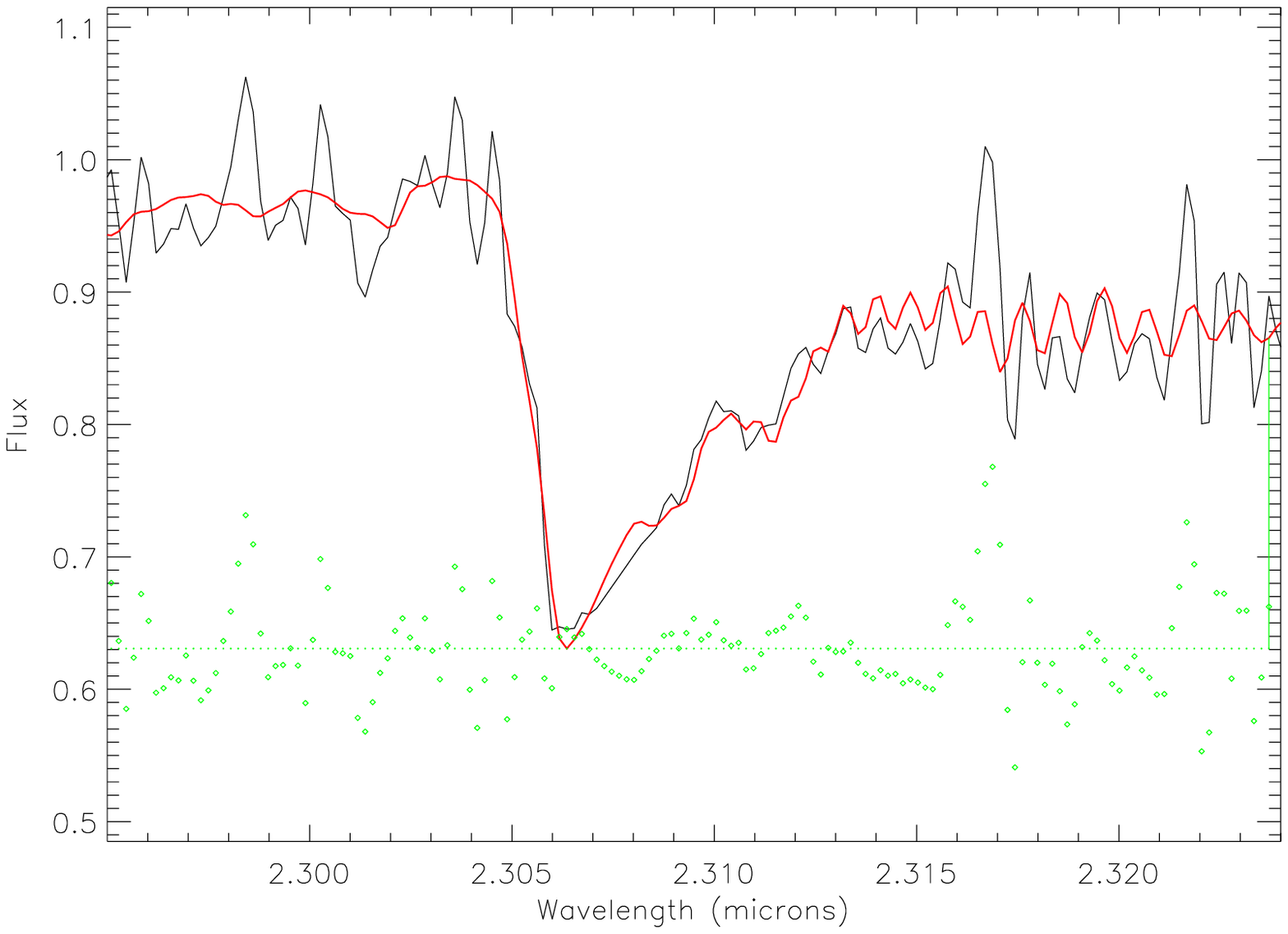}
\caption{Fits of the stellar kinematics for the positions  A
(top) and  B (bottom) shown in Figure\,\ref{large}. The observed
spectra are shown in black, the fitted template in red and the residuals in green.}
\label{fit}
\end{figure}

\begin{figure*}
\includegraphics[scale=1.15]{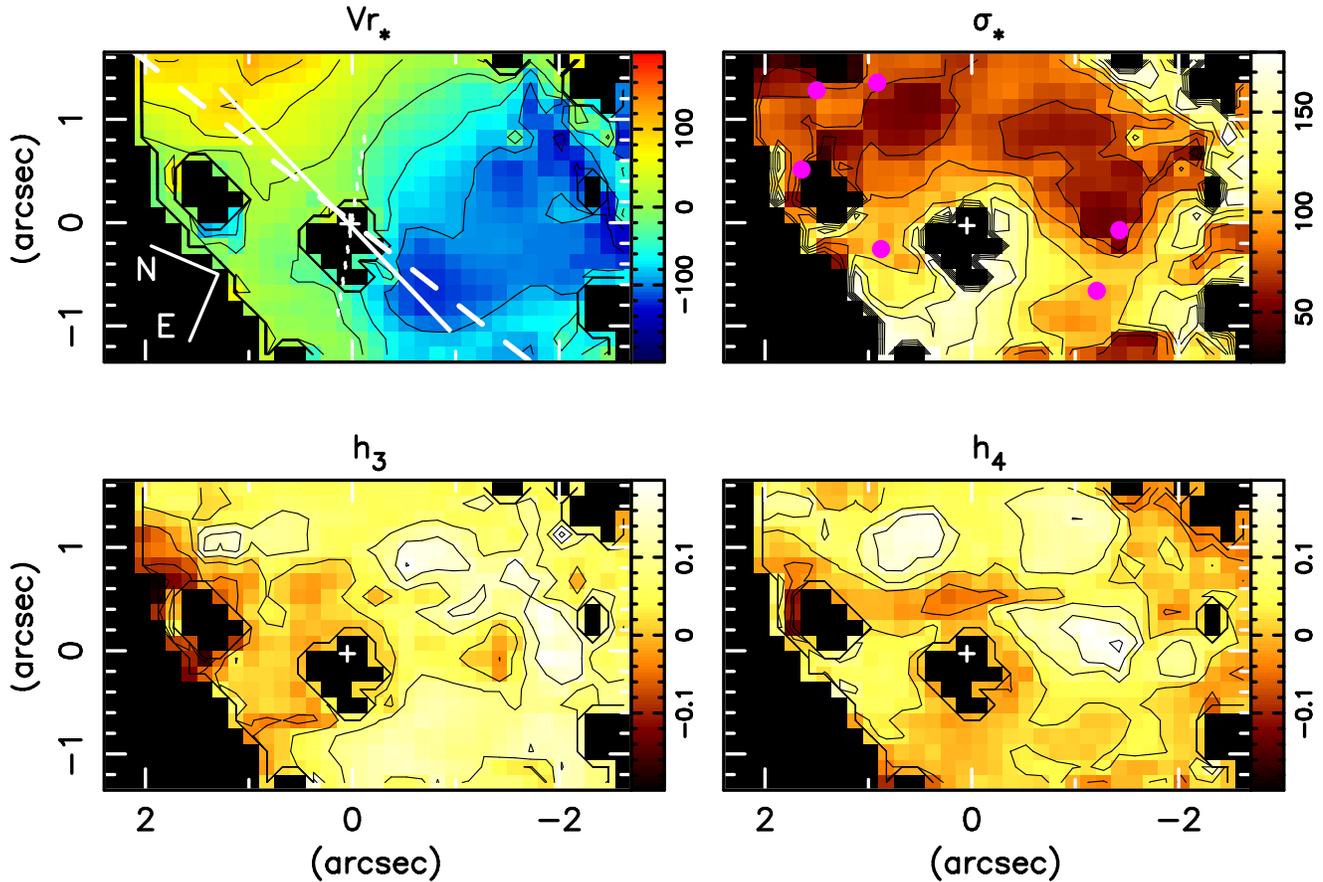}
\caption{2D maps of the stellar kinematics. Top left: radial velocities with
mean uncertainties of  15\,km\,s$^{-1}$. Top right: velocity
dispersions with uncertainties of 10\,km\,s$^{-1}$. Bottom: $h_3$ and
$h_4$ Gauss-Hermite moments with uncertainties of 0.03. The dashed
line represents the PA of the major axis of the galaxy, while the dotted and solid lines show the orientation and extent of the nuclear bar and nuclear spiral arms, respectively, observed by \citet{acosta-pulido03}. The cross marks the position of the nucleus and the magenta filled circles indicate the position of the star clusters  observed by \citet{wold06a}. The 
orientation is  shown in the top left panel.}
\label{stel}
\end{figure*}

In order to obtain the stellar line-of-sight velocity distribution (LOSVD) we fitted the $^{12}$CO(2,0) stellar 
absorption bandhead at 2.2935\,$\mu$m in the $K$-band spectra using the penalized
Pixel-Fitting (pPXF) method of \citet{cappellari04}. The algorithm
finds the best fit to a galaxy spectrum by convolving  template
stellar spectra with a given LOSVD. This procedure outputs the stellar radial
velocity, velocity dispersion  and higher-order Gauss-Hermite moments
for each spectrum fitted. The pPXF method allows the use of several
 stellar templates, and also to vary the relative
contribution of the different templates to obtain the best fit, thus
minimizing the problem of template mismatch. For this study, we have
selected as template spectra those of the Gemini spectroscopic library
of late spectral type stars observed with the GNIRS IFU using the
grating 111 l/mm in the {\it K}-band \citep{winge07}. In
\citet{riffel08}, we describe some properties of this library and
discuss the problem of template mismatch using the CO absorption
bandheads with the pPXF method.

In Figure\,\ref{fit} we show the resulting fit for
positions A (top) and B (bottom) marked in
Fig.\,\ref{large}. For the regions where the signal-to-noise
(hereafter S/N) ratio  was lower than $\approx30$ we replaced the
spectrum of each single pixel by an average  including the 9 nearest pixels. Even adopting this procedure, we could obtain measurements of the
stellar kinematics only for the central $4\farcs8\times3\farcs0$ region, as
the S/N ratio of the spectra close to the borders of the IFU
field becomes too low to obtain a reliable fit. For this same reason, we could not obtain stellar kinematic measurements also very close to the nucleus (radius$<$0\farcs3), as
the CO bandhead feature becomes diluted by dust and line emission.

In Figure\,\ref{stel} we present 2D maps of the stellar
kinematics. Black pixels in this figure correspond to the
masked regions where we did not obtain a reliable fit.
 The central cross marks the position
of the nucleus, defined as the locus of the peak of the $K-$band continuum
and the dashed line shows the PA=165$^\circ$, which is the position angle of the major axis of the galaxy \citep{morris85}.
 The top left panel of Fig.\,\ref{stel} shows the stellar radial
velocity field, from which we subtracted the systemic velocity of
the galaxy listed in the HyperLeda\footnote{http://leda.univ-lyon1.fr} database -- $V_s=1605\,{\rm km\,s^{-1}}$ 
(relative the local standard of rest), which agrees with our value for the nuclear region. The mean
uncertainty in the radial velocity is $\approx15\,{\rm km\,s^{-1}}$.  The radial
velocity field shows bluehisfts to the SE and redshifts to the NW, suggesting rotation, although with the kinematical centre displaced from the peak of the continuum emission.
The velocity field has an amplitude of $\approx 200\,{\rm km\,s^{-1}}$ (from  $-100\,{\rm km\,s^{-1}}$ to $100\,{\rm km\,s^{-1}}$) 
and seems not to be symmetric relative to the nucleus, with the highest blueshifts to the S--SW being reached closer to the nucleus than the highest redshifts to the N--NW. A one-dimensional cut along the galaxy major axis suggests that the turnover of the rotation curve to the S--SW is reached already at $\approx$100--150\,pc from the nucleus, but the turnover to the N--NW seems to be beyond the border of the field (see Fig.\,\ref{cut}).

In the top right panel of Fig.\,\ref{stel} we present the stellar
velocity dispersion ($\sigma_*$) map, with 
values ranging from 40 to 170\,km\,s$^{-1}$, and mean uncertainties of
$\approx10\,{\rm km\,s^{-1}}$. The highest values are observed to the E
and S of the nucleus. A
partial ring of low $\sigma_*$ values ($\approx50\,{\rm km\,s^{-1}}$)
is observed at $\approx 1^{\prime\prime}$ from the nucleus,  surrounding it from N to SW. 
The $h_3$ and $h_4$ Gauss-Hermite
moments maps, which measure deviations of the LOSVD from a Gaussian distribution,
are presented at the bottom panels of Fig.\,\ref{stel} and have mean
uncertainties of 0.03. The values of $h_3$ and $h_4$ are between
$-$0.18 and 0.18, which are similar to typical values obtained for a large number
of galaxies in similar analyses \citep{emsellem04,ganda06}.

\subsection{Gas kinematics}\label{res_kin}

\begin{figure*}
\includegraphics[scale=1.15]{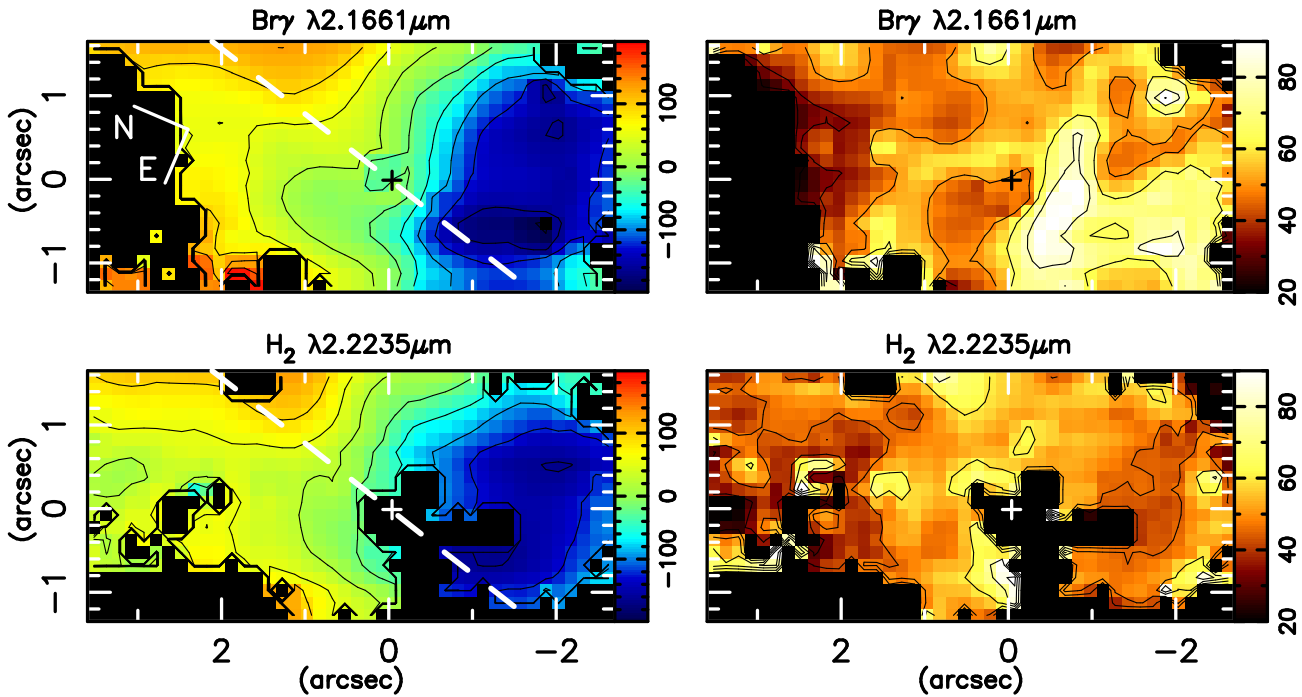}
\caption{Left: Radial velocity maps for the Br$\gamma$ (top) and
H$_2\,\lambda2.2235\,\mu$m (bottom) emitting gas, with mean
uncertainties of 6 and 9\,km\,s$^{-1}$, respectively. Right: Velocity
dispersion maps for the same emission lines with mean uncertainties of
7\,km\,s$^{-1}$ for Br$\gamma$, and 10\,km\,s$^{-1}$ for H$_2$.  The
dashed line represent the orientation of the major axis of the galaxy.}
\label{gas}
\end{figure*}

In the left panels of Figure\,\ref{gas} we present the radial velocity
fields obtained from the peak wavelengths of the Br$\gamma$ and
H$_2\,\lambda$2.2235$\,\mu$m emission lines, with mean uncertainties
of 6\,km\,s$^{-1}$ and 9\,km\,s$^{-1}$, respectively. The systemic
velocity of the galaxy has been subtracted from all the emission-line 
velocity maps. The velocity field of the Br$\gamma$ emitting gas is very similar to that of  the H$_2$ emitting gas, being also similar to the stellar velocity field, with the S--SE side approaching and the N--NW side receeding. 
The largest blueshifts  are observed S of the
nucleus reaching values of $\approx-170\,{\rm km\,s^{-1}}$, while the
largest redshifts reach smaller values of $\approx100\,{\rm km\,s^{-1}}$  and
are observed NW of the nucleus.

The gas velocity dispersion ($\sigma$) maps are shown in the right panels
of Fig.\,\ref{gas}. Mean uncertainties are  7\,km\,s$^{-1}$ for
Br$\gamma$ and 10\,km\,s$^{-1}$ for H$_2$.  The values 
were corrected for the instrumental broadening ($\sigma_{\rm inst}\approx18\,{\rm km\,s^{-1}}$), and range from $\sigma\approx20$ to $\approx90\,{\rm km\,s^{-1}}$ for Br$\gamma$ and  from $\sigma\approx20$ to $\approx70\,{\rm km\,s^{-1}}$ for H$_2$. The Br$\gamma$ velocity dispersion show regions of low values ($\sigma\approx30\,{\rm km\,s^{-1}}$) to the N of the nucleus, and
higher values to the S--SE. The highest $\sigma$ values observed in the
H$_2$ line are located around the nucleus, and the lowest values in
regions away from it. 

\subsection{Velocity Slices }



In order to better map the gaseous kinematics, and as our data has a relatively  high spectral  resolving power, we could ``slice''
the line profiles in a sequence of velocity channels. Our goal is to better sample the gas kinematics over the whole velocity distribution, including the emission-line wings
\citep[see][]{gerssen06,riffel06,riffel08}.   The flux distributions in the velocity slices were obtained after  subtraction of the continuum contribution determined as the average of the flux from both sides of the emission lines. Each slice corresponds
to a velocity bin of $\approx{\rm 50\,km\,s^{-1}}$ (two spectral
pixels) and the resulting flux distributions are shown in Figs.\,\ref{slices_br}
and \ref{slices_h2}. In these figures, the flux levels are presented in logarithmic units, and the slices trace the gas from negative (blueshifted, top left) to positive (redshifted, bottom right) velocities relative to the systemic velocity. For the Br$\gamma$ emitting gas, the highest blueshifts of $\approx-360\,{\rm km\,s^{-1}}$ are observed at
$\approx$1\arcsec\,SW of the nucleus, while the highest redshifts of
$\approx260\,{\rm km\,s^{-1}}$ are observed at $\approx2$\arcsec\,NW
of the nucleus. For the H$_2$ emitting gas, the highest blueshifts
reach $\approx-300\,{\rm km\,s^{-1}}$ and the highest redshifts
$\approx160\,{\rm km\,s^{-1}}$, observed aproximately at the same
locations as for the Br$\gamma$ emision. While the  highest blueshifts
are observed displaced from the major axis, in the velocity range
from $\approx-$200 to 200\,km\,s$^{-1}$ the emission peak moves from S--SE to N--NW essentially parallel to the major axis, suggesting that, for this range of
velocities, the emission is dominated by gas rotating in the plane of
the galaxy.

\begin{figure*}
\includegraphics[scale=1.15]{figs/slices_br.eps}
\caption{Velocity slices along the Br$\gamma$ emission line profile in
steps of $\approx50\,{\rm km\,s^{-1}}$. The dashed lines represent the orientation of the major axis of 
the galaxy and the velocity corresponding to the centre of each bin is shown
at the top left corner of the corresponding panel.}
\label{slices_br}
\end{figure*}

\begin{figure*}
\includegraphics[scale=1.15]{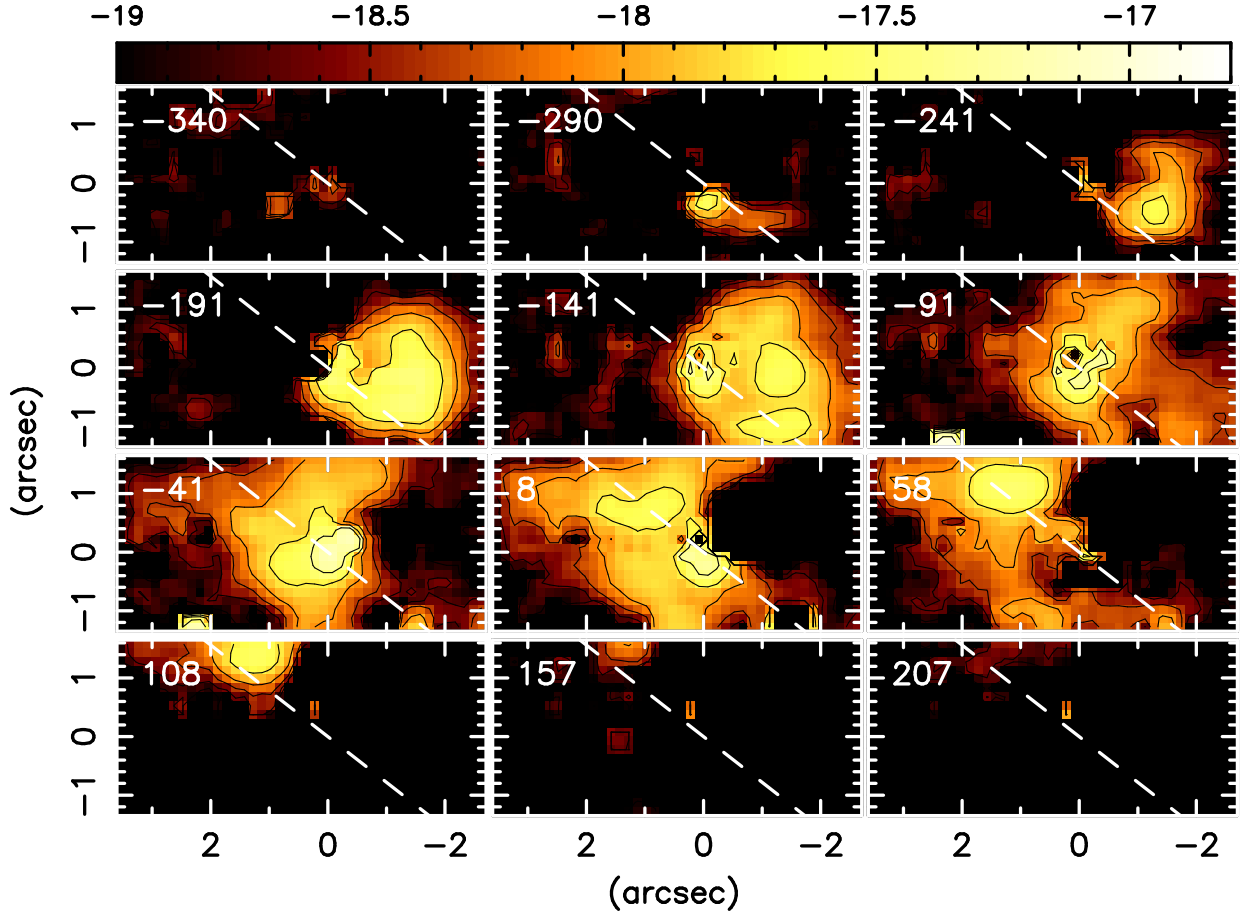}
\caption{Same as Fig.\,\ref{slices_br} for the H$_2\,\lambda$2.2235$\,\mu$m emission-line profile.}
\label{slices_h2}
\end{figure*}

\section{Discussion}

\subsection{The nuclear spectrum}
\label{nuclear}

\begin{figure}
\includegraphics[scale=0.4]{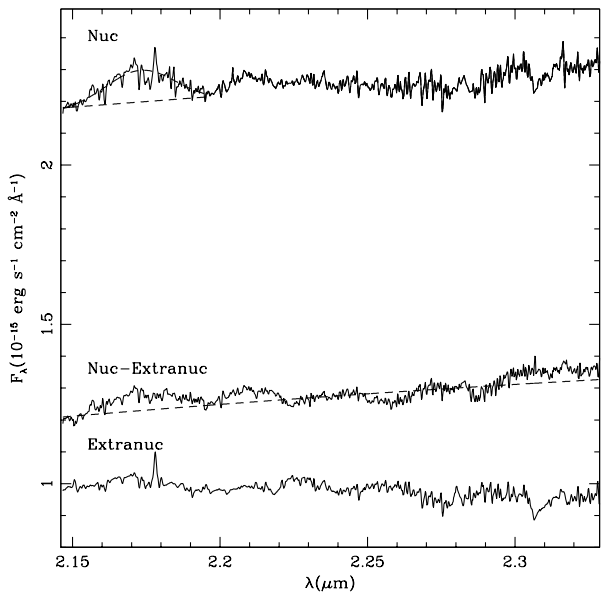}
\caption{Nuclear spectrum of NGC\,7582 for a circular aperture of 0$\farcs$8 diameter (top spectrum), extranuclear spectrum for a ring with inner radius of 0$\farcs4$ and outer radius of 0$\farcs6$ from the nucleus (bottom spectrum) and the difference between the two (middle spectrum). The extranuclear spectrum was scaled by the ratio between the areas of the nuclear and extranuclear apertures.}
\label{fitting}
\end{figure}

The flux distribution in the continuum shows an unresolved source at the nucleus (whose location we assume to coincide with the peak of the flux distribution). In order to isolate the spectrum of the nuclear source, we have extracted its spectrum within an aperture of radius 0$\farcs4$, which corresponds to the radial distance where the point source profile decreases to 1/3 of its peak value. In order to subtract the underlying stellar contribution to the nuclear flux we have extracted a spectrum within a ring with inner and outer radii of 0$\farcs$4 and  0$\farcs$6, respectively, surrounding the nuclear aperture. We show the apertures corresponding to the nuclear and extranuclear spectra as circles in the top panel of Fig.\,\ref{flux}. This choice of apertures was dictated by the need to isolate the nuclear spectrum from that of the circumnuclear ring of star formation. Larger apertures for the nuclear and extranuclear spectra would include contribution from  the star-forming ring. 

The resulting nuclear and extranuclear spectra are shown in Fig.\,\ref{fitting} (top and bottom spectra). In the same Figure, we also show the difference between them (middle spectrum) after scaling the circumnuclear spectrum to the flux it would have if obtained through the same aperture as that of the nuclear spectrum. We note that the circumnuclear spectrum does not differ much from those of the star-forming regions A and B, suggesting that recent star formation is occurring also inside the ring. This is supported by previous studies of the stellar population in the optical \citep{sb01} and near-infrared \citep{davies05}.


The nuclear spectrum  seems to present  both a narrow and a broad component in the Br$\gamma$ emission line. The fit of a Gaussian to the broad component, as illustrated in the top spectrum of Fig.\,\ref{fitting}, gives a FWHM\,$\approx$\,3900\, km\,s$^{-1}$, with the centre of the line blueshifted by $\approx650\,$km\,s$^{-1}$ relative to the narrow component. The presence of a broad component in Br$\gamma$, with FWHM  $\approx$4000 km\,s$^{-1}$  has been previously reported by \citet{davies05}, in long-slit observations obtained with the Very Large Telescope. A broad component, observed in H$\alpha$,  was  also reported by \citet{aretxaga99}, although this component seems  to have been produced by a transient event. These authors noted the appearance and subsequent fading of the broad H$\alpha$ component as well as other optical emission lines and proposed three possible scenarios for its origin: capture and disruption of a star by a SMBH, a reddening change in a dusty torus surrounding the nucleus, and a Type IIn supernova exploding in a compact starburst. 

The nuclear spectrum is redder than the extranuclear one, and the subtraction of the stellar contribution does seem to have eliminated the stellar 2.3$\mu$m CO absorption, clearly observed prior to the subtraction. The resulting spectrum shows a red continuum similar to those of other red unresolved components observed in other AGN and attributed to emission from hot dust, possibly distributed in a toroidal structure around the SMBH, with temperature close to the dust sublimation temperature  \citep{barvainis87,marco98,marco00,rodriguez-ardila05b,rodriguez-ardila06}. Under the assumption that the red unresolved component is indeed due to emission by dust, we have derived its temperature and mass, by fitting a blackbody function to its spectrum.  The resulting fit is shown as a dashed line in the middle spectrum of Fig.\,\ref{fitting}, corresponding to a temperature $T\approx1050\pm140$\,K, for which the maximuum of the blackbody curve  is reached  at $\approx2.76\,\mu$m.

We have estimated the mass of the hot dust producing the emission in the unresolved component using the formalism of \citet{barvainis87}, for  dust composed by grains of graphite. The infrared spectral luminosity of each dust grain, in erg\,s$^{-1}$\,Hz$^{-1}$, can be written as:
\begin{equation}
 L^{\rm gr}_{\nu,{\rm ir}}=4\pi^2 a^2 Q_\nu B_\nu(T_{\rm gr}),
\end{equation}
where $a$ is the grain radius, $Q_\nu$ is its absorption efficiency and $B_\nu(T_{\rm gr})$ is its spectral distribution assumed to be a Planck function for a temperature $T_{\rm gr}$.

The total number of graphite grains can be obtained from:

\begin{equation}
N_{\rm HD}\approx \frac{L^{\rm HD}_{\rm ir}}{L^{\rm gr}_{\rm ir}}
\end{equation}

\noindent where $L^{\rm HD}_{\rm ir}$  is the total luminosity of the hot dust, obtained by integrating the flux under the function fitted to the nuclear--extranuclear spectrum
and adopting a distance to NGC\,7582 of $d=21.6\,$Mpc. In order to obtain $L^{\rm gr}_{\rm ir}$ we have adopted $T_{\rm gr}=1050$\,K,  as derived above, $a=0.05\,\mu$m and $Q_\nu=1.4\times10^{-24}\nu^{1.6}$ (for $\nu$ in Hz) as in \citet{barvainis87}.
Finally, we can obtain the total mass of the emitting dust as:

\begin{equation}
M_{\rm HD}\approx\frac{4\pi}{3} a^3 N_{\rm HD}\rho_{\rm gr}
\end{equation}

Adopting $\rho_{\rm gr}=2.26\,$g\,cm$^{-3}$ for the graphite density \citep{granato94}, we obtain $M_{\rm HD}\approx2.8\times10^{-3}\,{\rm M_\odot}$. In Table\,\ref{masses} we compare this dust mass with masses of hot dust obtained for other AGN in previous studies. The hot dust mass in NGC\,7582 is similar to those obtained for Mrk\,766, NGC\,1068 and NGC\,3783, while it is smaller than the values for Mrk\,1239, NGC\,7469 and Fairall\,9, and higher than those derived for NGC\,1566 and NGC\,4593.


\begin{table*}
\centering
\caption{Masses of hot dust found in AGNs}
\vspace{0.3cm}
\begin{tabular}{l l l}
\hline
 Galaxy                & $M_{\rm HD}$ (M$_\odot$)& Reference \\
\hline
NGC\,7582 &              $2.8\times10^{-3}$          & This work                           \\
Mrk\,1239 &              $2.7\times10^{-2}$          &  \citet{rodriguez-ardila06}           \\
Mrk\,766   &             $2.1\times10^{-3}$          & \citet{rodriguez-ardila05b}          \\
NGC\,1068 &              $1.1\times10^{-3}$          &  \citet{marco00}                     \\
NGC\,7469 &              $5.2\times10^{-2}$          & \citet{marco98}                      \\
NGC\,4593  &             $5.0\times10^{-4}$          & \citet{santos95}                      \\ 
NGC\,3783  &             $2.5\times10^{-3}$          & \citet{glass92}                      \\
NGC\,1566  &             $7.0\times10^{-4}$          &  \citet{baribaud92}                  \\
Fairall\,9 &             $2.0\times10^{-2}$          & \citet{clavel89}                     \\
\hline
\end{tabular}
\label{masses}
\end{table*}

For Mrk\,1239 \citep{rodriguez-ardila06} and Mrk\,766 \citep{rodriguez-ardila05b} the dust mass values were derived from long-slit spectroscopy, with extraction apertures of the nuclear spectrum  corresponding to more than 100\,pc at the galaxies, while the spatial resolution of our data allow us to constrain the location of the hot dust as being within $\approx 25$\,pc from the nucleus. As pointed out above, the $K-$band excess is probably due to the emission of hot dust from a circumnuclear torus which may be located even
closer to the SMBH. In the case of NGC\,1097, for example, \citet{sb05} have found a heavily obscured starburst closer than 9\,pc from the nucleus, which could be located in the outskirts of the torus. In the case of NGC\,1068, interferometric mid-IR observations constrained the hot dust emission to originate much closer, in a region of  aproximately 1\,pc diameter \citep{jaffe04}. 

\subsection{Circumnuclear star-forming regions}

NGC\,7582 is known to present a kpc-scale H$\alpha$ emitting disk which harbors recent star formation plus heavy dust obscuration in the nuclear region
\citep{morris85,regan99,sb01,sosa-brito01,wold06a}.  The presence of dust to the NE is responsible for the assymetric contours of the continuum flux map in the bottom panel of Fig.\,\ref{flux}. The recent star-formation in the dust-embedded clusters reported by \citet{wold06a} (shown as magenta circles in Fig. \,\ref{flux}) are seen as enhancements in our Br$\gamma$ image, delineating a ring of major axis of $\approx$3$\farcs$5 and minor axis of $\approx$1$\farcs$5. This gives an axial ratio of $\frac{a}{b}\approx2.3$, which is similar to the axial ratio of the galaxy ($\frac{a}{b}\approx2.2$\footnote{from HyperLeda database (http://leda.univ-lyon1.fr)}), suggesting that the star-forming ring is circular, with a radius of $\approx$190\,pc  and is located in the plane of the galaxy. As pointed out by \citet{wold06a}, these star-forming clusters had not been previously seen neither in the optical nor in the near-infrared. The near-IR IFU data from \citet{sosa-brito01}, although consistent with our results, does not have the necessary spatial resolution to resolve these clusters.

In our data alone, clear peaks in the Br$\gamma$ flux distribution are  located in the ring at $1\farcs7$\,NW and $1\farcs7$\,SE of the nucleus, and are encircled by the green contours in Fig.\,\ref{flux}, which comprise the pixels with  fluxes higher than 50\% of the peak flux. We will refer to each of these two regions -- whose physical boundaries we define as corresponding to the green contours  -- as  circumnuclear star-forming regions (CNSFRs), and will use our data for these regions  to characterize the star-formation in the circumnuclear ring. From the locations of the star clusters found by \citet{wold06a} it can be seen that each of these CNSFRS correspond to two star clusters, thus we should bear in mind that the parameters we obtain for these regions correspond to two star clusters instead of one.


Following \citet{riffel08}, we can estimate the mass of ionized hydrogen in each CNSFR as: 

\begin{equation}
 \left(\frac{M_{\rm HII}}{\rm
 M_\odot}\right)=2.88\times10^{17}\left(\frac{F_{\rm Br\gamma}}{\rm
 erg\,s^{-1}cm^{-2}}\right)\left(\frac{d}{\rm Mpc}\right)^2, 
\label{mhii}
\end{equation}
where $F_{Br\gamma}$ is the line flux, $d$ is the distance of the galaxy
and we have assumed an electron temperature $T=10^4$\,K and
electron density $N_e=100\,{\rm cm^{-3}}$ for the CNSFRs. 



We can also obtain the  emission rate of ionizing photons  for each CNSFR using \citep{osterbrock89}:
\begin{equation}
Q[H^+]=\frac{\alpha_B\,L_{H\alpha}}{\alpha^{\rm eff}_{H\alpha}\,h\nu_{H\alpha}},
\end{equation}
where $\alpha_B$ is the hydrogen recombination coefficient to all energy levels  above the ground level, $\alpha^{\rm eff}_{H\alpha}$ is the effective recombination coefficient for $H\alpha$, $h$ is the Planck's constant and  $\nu_{H\alpha}$ is the frequency of the $H\alpha$ line. Using $\alpha_B=2.59\times10^{-13}\,{\rm cm^3s^{-1}}$, $\alpha^{\rm eff}_{H\alpha}=1.17\times10^{-17}\,{\rm cm^3s^{-1}}$ and $L_{H\alpha}/L_{Br\gamma} = 102.05 $ for  case B recombination assuming an electronic temperature $T_e=10^4\,{\rm K}$  \citep{osterbrock89}, we obtain:   
\begin{equation}
\left(\frac{Q[H^+]}{\rm s^{-1}}\right)=7.47\times10^{13}\,\left(\frac{L_{Br\gamma}}{\rm erg\,s^{-1}}\right).
\label{q}
\end{equation}

Finally, we can also obtain the star formation rate (SFR) using \citep{kennicutt98}:
\begin{equation}
\left(\frac{SFR}{\rm M_\odot yr^{-1}}\right)=8.2\times10^{-40}\,\left(\frac{L_{Br\gamma}}{\rm erg\,s^{-1}}\right),
\end{equation}
under the same asumptions adoped to derive equation\,\ref{q}.


As there is considerable obscuration in the nuclear region of NGC\,7582, the emission-line fluxes used in the above equations have been corrected for reddening. We have
used $E(B-V)=0.6$, obtained by \citet{schmitt99} in a spectrum extracted withing an aperture of 2$\times$2\,arcsec$^2$, adopting  the extinction  law of \citet{cardelli89}. 

Besides calculating the above quantities for the two CNSFRs, we have also calculated them for the nucleus and for the complete ring of star formation, as follows. In order to obtain the emission rate of ionizing photons for the nucleus, we integrated the Br$\gamma$ emission-line flux within a circular aperture of 0$\farcs$4 plus that emitted by  the outflowing gas, under the assumption that the outflowing gas is ionized by the active nucleus. The flux in the outflowing gas was estimated as the sum of the fluxes in the velocity slices with $|v|>200$\,km\,s$^{-1}$. In this manner, we obtain only  a lower limit for the rate of ionizing photons emitted by the nucleus, as our field of view covers only a very small portion of the outflow, as discussed in Sec.\,\ref{results}. An upper limit can be obtained under the assumption that the total flux in the outflow is the one observed times the ratio between the total area covered by the outflow (estimated from the [O\,{\sc iii}] image) and that covered by our observations, which is $\approx$5. Considering that there should be a counterpart outflow to the opposite side of the galaxy plane, probably hidden by the dust in the plane, the Br$\gamma$ flux  in the outflow may be $\approx$10 times the one measured from our data.
The Br$\gamma$ flux of the star-forming ring was obtained by integrating its value over the whole IFU field and  subtracting the contribution of the nuclear flux plus that of the outflowing gas.
 
Table\,\ref{par} presents the results of the above calculations, together with the area used in the integrations and the Br$\gamma$  emission-line luminosities corrected for reddening, For the nucleus we list both the lower limit (including only the contribution of the observed outflow) and the upper limit (including the contribution of the total estimated outflow). We also list
the equivalent widths of the Br$\gamma$ emission line.

\begin{table*}
\centering
\caption{Physical parameters obtained for the CNSFRs in NGC\,7582.}
\vspace{0.3cm}
\begin{tabular}{l c c c c}
\hline
 Parameter                              & $1\farcs7$\,NW & $1\farcs7$\,SE & Nucleus$^{\rm(nuc+ulo)^a}_{\rm (nuc+llo)^b} $ & Ring \\
\hline
Area (arcsec$^2$)                       &  1.48            & 1.15             &  0.5   &   18.4   \\
log\,$L({\rm Br\gamma})$ [erg\,s$^{-1}$]&  38.53$\pm$0.03  & 38.45$\pm$0.03   &  37.81$_{(37.96)}^{(38.52)}\pm$0.04 &   39.09$\pm$0.03  \\
log\,$Q[H^+] {\rm [s^{-1})]}$           &  52.40$\pm$0.03  & 52.32$\pm$0.03   & 51.68$_{(51.83)}^{(52.39)}\pm$0.04  &    52.96$\pm$0.03   \\
$SFR {\rm (M_\odot yr^{-1}})$           & 0.28$\pm$0.02    & 0.23$\pm$0.02    &  --             &    1.01$\pm$0.07   \\
EW(Br$\gamma$) [\AA]                    &  14.7$\pm$1.4    &  13.6$\pm$1.3    & 0.8$\pm$0.1     & 7.1$\pm$0.6       \\
$M_{\rm HII}$ ($10^5$ M$_\odot$)        &  8.2$\pm$0.6       & 6.8$\pm$0.5    & 1.5$_{(2.2)}^{(8.2)}\pm$0.2     &   29.4$\pm$2.0      \\
\hline
\multicolumn{5}{l}{$^a$ Nucleus plus upper limit for the outflowing gas component.}\\
\multicolumn{5}{l}{$^b$ Nucleus plus lower limit for the outflowing gas component.}\\
\end{tabular}
\label{par}
\end{table*}

Table\,\ref{par} shows that the Br$\gamma$ flux in inner $\approx$400\,pc of NGC\,7582  is dominated by emission from the circumnuclear star-forming ring, which  ranges from  4 to 13 times that of the nucleus plus outflow. The Br$\gamma$ fluxes of the 
two CNSFRs contribute with  50\,\% of the total emission from the star formation ring.

The emission rates of ionizing photons obtained for the two CNSFRs are in good agreement with values obtained for CNSFRs in other galaxies \citep[e.g.][]{galliano08,hagele07}, as well as with rates derived for NGC\,7582 in the Mid-IR \citep{wold06a}. Such rates correspond to about 1000 O6 stars in each CNSFR \citep{osterbrock89}. Considering that each CNSFR actually corresponds to two clusters, then this would correspond to $\approx$500 O stars per cluster.

The star-formation rates for the CNSFRs of $SFR\approx0.23-0.28\,{\rm M_\odot/yr^{-1}}$, caracterize a moderate star-forming regime, and are in good agreement with the values derived by \citet{shi06} for a sample of 385 CNSFRs in galaxies covering a range of Hubble types.

\begin{figure}
\includegraphics[angle=-90,scale=0.6]{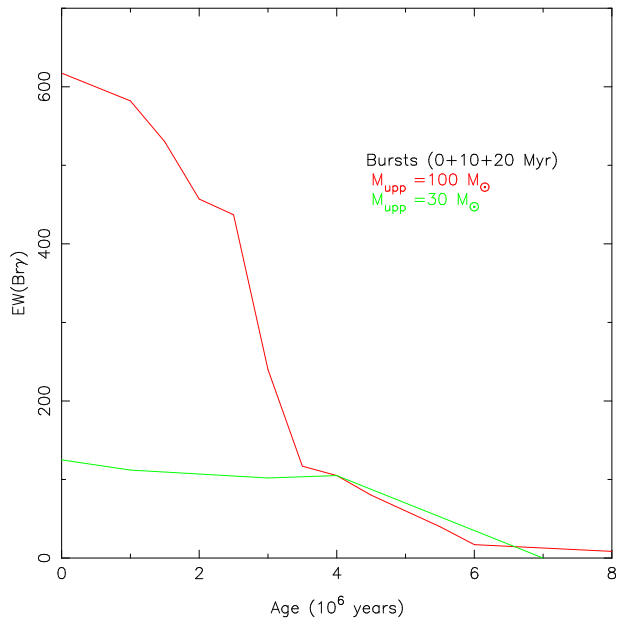}
\caption{Equivalent width of Br$\gamma$ as predicted by evolutionary
photoionization models \citep{dors08}.}
\label{photo}
\end{figure}

\subsubsection{Age of the CNSFRs}

We can use our data also to estimate the age of the CNSFRs, applying
the method described by \citet{dors08}, which consists of comparing the observed
values of EW(Br$\gamma$) to those predicted by evolutionary
photoionization models. As stressed by those authors, the real age of
the CNSFRs can only be obtained if the contributions both from the
underlying bulge and from older stars in the CNSFRs are taken into account.

In order to obtain the contribution of the underlying bulge, we used aperture
photometry in the continuum image (Fig.\,\ref{flux}) to measure both the flux of the 
CNSFRs continua and that of the bulge in the surrounding regions. We
find a bulge contribution of $\approx$50 \% of the total flux at both  CNSFRs, and thus the  value of EW(Br$\gamma$) increases from 15 \AA~ to 30 \AA\ after this correction.

The presence in the CNSFRs of stars formed in previous bursts of star formation is supported by the work of \citet{schmitt99}, who found a large spread of
age in the circumnuclear star-forming regions. In order to 
take this effect into account, we apply the same approach as
\citet{dors08}, representing the CNSFR stellar content by three bursts
of star formation, with ages of 0.1, 10, and 20 Myr, and considering two values
for the upper limit in stellar mass ($M_{\rm up}$= 100 and 30
$M_{\odot}$). The resulting evolutionary behaviour of
EW(Br$\gamma$) is illustrated in Figure~\ref{photo}. 
For a value of EW(Br$\gamma$)$\approx$30\AA, we obtain an age of
$\approx$5.0 Myr, which is similar to the values found for the CNSFRs in NGC\,6951 and
NGC\,1097 by \citet{dors08}.

\subsection{Stellar Kinematics} \label{stelkinematics}

As observed in Fig.\,\ref{stel} the stellar velocity field seems to be dominated by rotation. We have thus tried to fit the stellar velocities by a model 
of circular orbits in a plane subject to  a Plummer potential, which sucessfully reproduced the stellar kinematics of the central regions of other Seyfert galaxies in previous studies \citep[e.g.][]{barbosa06,riffel08}. Nevertheless,  in the case of NGC\,7582 this model fails, as the residuals are too large everywhere. The kinematic centre of the rotation in Fig.\ref{stel} seems to be displaced from the nucleus by $\approx0\farcs$5\,NW ($\approx$50\,pc at the galaxy), an effect which can also be observed in one-dimensional cuts of the velocity field shown in Fig.\,\ref{cut}. This is an interesting result which should be further investigated. Could this be the kinematic centre of the galaxy? If this is the case, then the peak flux of the infrared continuum would not correspond to the centre of the galaxy but instead to the location of the SMBH (the unresolved red source), and the ring of star formation would be centred on the SMBH instead of the galaxy centre. We believe this is unlikely, as the radius of influence of the SMBH is much smaller than the radius of the ring, and thus one would expect that the ring would be centred on the galaxy nucleus. Our field of view is too small for us to reach a firm conclusion, but one possibility is that the apparent displacement of the centre of the rotation field can be due to distortions in a circular velocity field.  Observations in the mid-IR  \citep{acosta-pulido03} show that NGC\,7582  presents both a nuclear bar and two nuclear spiral arms, which could be related to the observed distortions. The nuclear bar is oriented along PA\,$\approx110-290(-70)^\circ$ extending from $\approx 1^{\prime\prime}$\,E  to $\approx 1^{\prime\prime}$\,W from the nucleus, while the nuclear spiral arms are oriented approximately along the major axis of the galaxy and extending to $\approx1\farcs5$\,NW and SE of the nucleus. We show the orientation and extents of the nuclear bar and spiral as dotted and solid white lines, respectively in Fig.\,\ref{stel}. We note that the SE arm seems to correspond to the location where there seems to be an ``excess blueshift" in the stellar radial velocity field in Fig.\,\ref{stel}.


The stellar velocity dispersion shows the highest values
of $\approx$170\,km\,s$^{-1}$ to E and S of the nucleus (Fig.\,\ref{stel}) 
which we interpret as representing the $\sigma_*$
of the galaxy bulge. We use this value to estimate the mass of the SMBH
($M_{\rm BH}$) of NGC\,7582 from the relation ${\rm log}(M_{BH}/{\rm
M_\odot})=\alpha+\beta\,{\rm log}(\sigma_*/\sigma_0)$, where
$\alpha=8.13\pm0.06$, $\beta=4.02\pm0.32$ and
$\sigma_0=200$\,km\,s$^{-1}$ \citep{tremaine02}. For
$\sigma_*\approx170\,{\rm km\,s^{-1}}$, we obtain $M_{\rm
BH}\approx7\times10^7\,{\rm M_\odot}$, in good agreement with the
value obtained by \citet{wold06b} through modelling of the kinematics of [Ne\,{\sc
ii}]12.8$\mu$m emitting gas.

The $\sigma_*$ map  (Fig.\ref{stel}) shows a region  with low values ($\sigma_*\approx50\,{\rm km\,s^{-1}}$) delineating a partial ring around the nucleus and which seems to be embedded in  the bulge.  These locations are close to those of the star clusters observed by \citet{wold06a}, but somewhat displaced, being $\approx$\,50\,pc closer to the nucleus. The lower $\sigma_*$ values reveal that these stars have a ``colder kinematics'' than that of the stars of the bulge. We note that these $\sigma_*$ values are similar to those of the H$_2$ gas, suggesting that these stars have recently formed  in the circumnuclear ring from kinematically cold gas, and which have not yet ``thermalized'' with the ``hotter'' stars in the bulge. But in order to present CO absorption, the stars should be at least 10$^7$\,yr old \citep{oliva95}, what suggests that they belong to a previous burst of star-formation and not the one which is presently ionizing the gas. This interpretation is supported by the displacement of this ring relative to that in the ionized gas, which suggests that the star-formation is propagating outwards from the nucleus. Ring-like regions with low $\sigma_*$ values have been observed also around other AGN \citep{barbosa06,riffel08}.



\subsection{Gas Kinematics} \label{kinematics}

\begin{figure}
\includegraphics[scale=0.42]{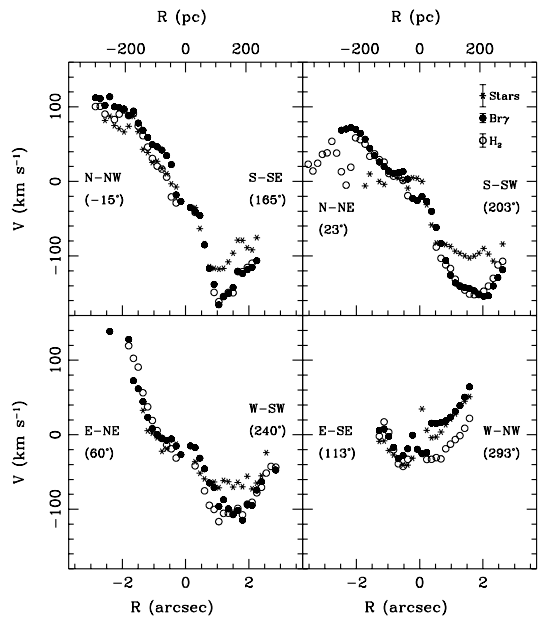}
\caption{One-dimensional cuts of the gaseous (black circles and
crosses) and stellar (open circles) radial velocities obtained along the PAs $-15-165^\circ$, $23-203^\circ$, $60-240^\circ$ and $113-293^\circ$ for a pseudo-slit of $0\farcs45$ width.}
\label{cut}
\end{figure}

The radial velocity maps of both the Br$\gamma$ and H$_2$ emitting gas (Fig. \ref{gas}) are similar to the stellar one (Fig. \ref{stel}). In order to better observe the possible differences, we have extracted one-dimensional cuts from the radial velocities maps which are shown in Figure\,\ref{cut}. These cuts were obtained by averaging the velocities within  pseudo-slits of $0\farcs45$ width passing through the nucleus. The orientations of the slits were selected to provide a good coverage of the velocity fields, and are shown as dashed green lines in the bottom panel of Fig.\,\ref{flux}.
The top-left panel of Fig.\,\ref{cut} shows the velocities along  PA=$-15(345)-165^\circ$, which is the orientation of the major axis of the galaxy \citep{morris85}, while in the top-right panel we present the velocities along PA$=23-203\degr$, in the bottom-left panel along PA$=60-240\degr$ and in  the bottom-right panel  along PA=$113-293\degr$.

The highest differences between the gaseous and stellar velocities are observed at 0$\farcs$5$-$2$\farcs5$\,SW from the nucleus along PA$=203\degr$,
where the gaseous velocities reach values of up to $-160\,{\rm km\,s^{-1}}$, while the stellar velocities reach at most $-100\,{\rm km\,s^{-1}}$. For the PAs $165\degr$ and 240$\degr$ the gaseous velocities are still more blueshifted than the stellar but only by about 40\,km\,s$^{-1}$ (to SE--SW of the nucleus).  Along PA=$293(-67)\degr$, to NW--W of the nucleus, the velocities observed for the stars are similar to those of the Br$\gamma$ emitting gas, while the H$_2$ emitting gas is blueshifted by $\approx-$20\,km\,s$^{-1}$. We note that this orientation coincides with that of the nuclear bar, and suggest that, as this blueshift is observed against the far side of the galactic plane, we could be observing an inflow of molecular gas towards the nucleus along the bar (which could be in the galaxy plane). We have already observed similar H$_2$ inflows towards the nucleus of another  active galaxy, NGC\,4051, not along a nuclear bar, but along a nuclear spiral arm  \citep{riffel08}.

From Fig.\,\ref{cut} it can be concluded that the emitting gas is blueshifted relative to the stars to SW of the nucleus. This region is co-spatial with part of  the [O\,{\sc iii}] ionization cone observed by \citet{sb91}, shown by the green contours in Fig.\,\ref{large}. We interpret this excess blueshift as due to the contribution  of gas outflowing from the nucleus to the total gas emission.  We note that we are observing only a small fraction of the outflowing gas because the field-of-view of our observations is much smaller than the dimensions of the ionization cone, along which the gas is outflowing \citep{morris85}.

\begin{figure}
\includegraphics[scale=1.1]{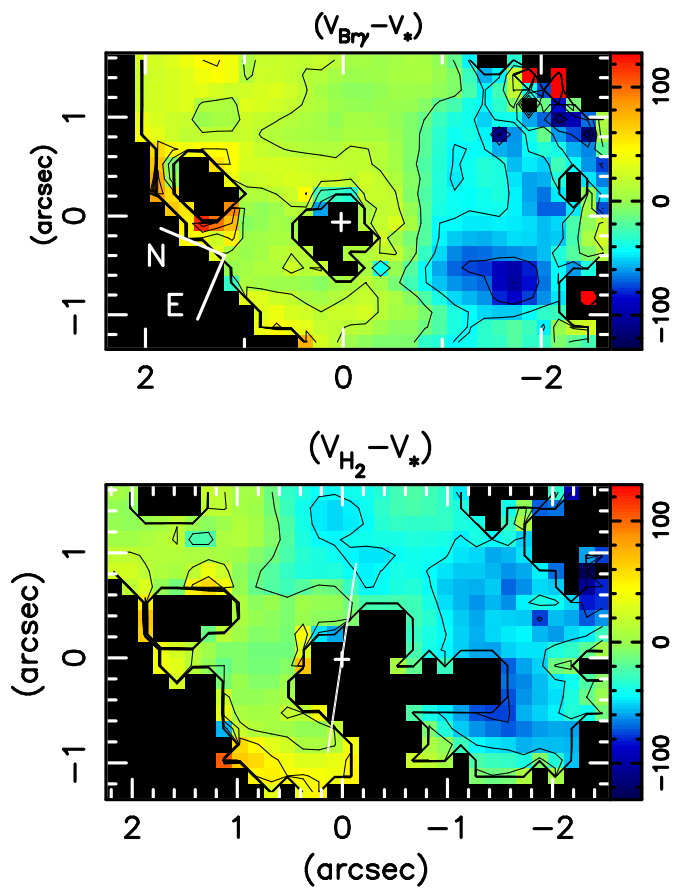}
\caption{Residual maps obtained by the difference between the
Br$\gamma$ and H$_2$ radial velocity fields and the stellar one. The white line in the bottom panel show the orientation and extent of the nuclear bar observed by \citet{acosta-pulido03}.}
\label{res}
\end{figure}

Figs. \ref{stel}, \ref{gas} and \ref{cut} suggest that the radial velocity field of the gas is similar to that of the stars except for the region along the ionization cone/nuclear outflow and possibly along the nuclear bar where there may be an H$_2$ inflow. In order to isolate the outflowing and inflowing components, we have subtracted the stellar velocity field from the Br$\gamma$ and H$_2$ velocity fields. These residual maps are shown 
in Figure\,\ref{res}. In the residual Br$\gamma$ velocity map the velocity values to  NE--N--NW of the nucleus are close to zero, while to the S--SW--W the gas is blueshifted by up to $-100$\,km\,s$^{-1}$ relative to the stars. This velocity is of the order of those found by \citet{morris85} for the [O\,{\sc iii}] emitting gas. In the residual H$_2$ velocity map, the blueshift at PA\,$\approx$\,293(-67) can be observed in approximate alignment with the nuclear bar, whose orientation is shown as the white solid line in the bottom panel of Fig.\,\ref{res}.

Even higher blueshifts -- of up to $-350\,{\rm km\,s^{-1}}$ -- are observed to the S--SW in the velocity channel maps (Fig.\,\ref{slices_br}). The difference between the radial velocity maps and  the velocity channel maps is due to the fact that the first give the peak velocity of the emission line for a given spatial position, while the latter show how the gas with a
certain velocity is distributed along the NLR. The peak wavelength will usually correspond to the 
kinematics of the component(s) with the highest flux(es), while the channel maps will reveal also weaker components which contribute to the wings of the emission-line profiles.
We thus attribute the high blueshifts seen in the velocity channel maps
to low mass, high velocity outflows, originated close to the nucleus which compress and push  the gas further out to the observed radial velocities of $\approx-$100\,km\,s$^{-1}$. These high velocity outflows do not appear in the radial velocity maps due to their low flux, appearing only as a blue wing to the profile, thus not affecting the peak wavelenght used to derive the radial velocity. We have observed  similar results in a recent study of the NLR of other nearby Seyfert galaxies  using IFUs \citep{barbosa09}.

Outflows from the nucleus of active galaxies have been observed at
optical wavelengths using long-slit spectroscopy \citep[e.g.][]{sb92,crenshaw00,das05,das06}
and from radio observations \citep[e.g.][]{morganti07} and more recently by our group
using IFUs \citep{riffel06,barbosa09}.



\subsubsection{Mass outflow rates}

 We have used the velocity channel maps of Fig.\,\ref{slices_br}
in order to estimate the mass outflow rate in ionized gas. Inspecting these maps, it can be concluded that radial velocities of up to $\pm$200\,km\,s$^{-1}$ are observed to both sides fo the nucleus, suggesting that up to these velocities, we are mapping gas motions in the galaxy disk. Above 200\,km\,s$^{-1}$, there are mostly blueshifts and almost no redshifts, what can be interpreted as due to the presence of ouflows. Our interpretation for this results is that these outflows are seen in blueshift to the S--SW because they are partially oriented  towards us and are projected against the far side of the galaxy disk while the  counterpart redshifted outflows are not seen because they are hidden behind the near side of the dusty disk. Thus, assuming that the emitting gas with blueshifts higher than 200\,km\,s$^{-1}$ is outflowing from the nucleus, we can estimate the mass outflow rate. In order to do this, we calculate the mass flux through a circular cross section with radius $r\approx0\farcs7$ at $1\farcs2$ from the nucleus, estimated from the shape of  the light distribution in the velocity channels corresponding to $-308$ and $-257\,{\rm km\,s^{-1}}$ in Fig.\,\ref{slices_br}. The mass outflow rate has been calculated as $\dot{M}_{HII}= 2 m_p\,N_e\,v\,\pi r^2\,f$, where $m_p$ is the proton mass, $N_e$ is the electron density, $v$ is the velocity of the outflowing gas and $f$ is the filling factor \citep{sb07,riffel08} and the factor of 2 is to take into account the likely presence of a hidden counterpart outflow to the one we observe. Adopting $N_e=100\,{\rm cm^{-3}}$ and $f=0.001$ and adding the contribution of the two velocity bins above, we obtain $\dot{M}_{HII}\approx4.8\times10^{-2}\,{\rm M_\odot\,yr}$.

This outflow rate is somewhat smaller than  those  quoted in the literature for active galaxies, which are between 0.1 and 10\,${\rm
M_\odot\,yr^{-1}}$ \citep[e.g.][]{veilleux05}. However, the latter values are
generally obtained for objects with higher levels of activity than
NGC\,7582. For similar levels of activity and in comparable scales of hundreds of parsecs, \citet{barbosa09} have obtained similar ionized gas outflow rates of
$\approx10^{-3}-10^{-2}\,{\rm M_\odot\,yr^{-1}}$ from 2D mapping of the gas
kinematics in the central region of 6 nearby Seyfert galaxies, in good agreement with the value obtained here.

We now calculate the mass accretion rate necessary to produce the observed nuclear luminosity of NGC\,7582. The latter can be estimated from $\dot{m}=\frac{L_{\rm bol}}{c^2\eta}$, where $L_{\rm bol}$ is the bolometric luminosity, $\eta$ is the efficiency of conversion of the rest mass energy of the accreted material into radiation and $c$ is the speed of light. We adopt $\eta=0.1$, a typical value for  geometrically thin and optically thick accretion discs \citep[e.g.][]{frank02} and use the relation $L_{\rm bol}\approx20\,L_{\rm X}$ \citep{elvis94}, 
where $L_{\rm X}\approx7.2\times10^{41}\,{\rm erg\,s^{-1}}$ is the hard X-ray 
luminosity \citep{piconcelli07}. The result is  $\dot{m}\approx2.6\times10^{-3}\,{\rm M_\odot\,yr^{-1}}$.  

Comparing the mass outflow rate with the accretion rate calculated above, it can be concluded that the former is an order of magnitude larger than the latter, a relation similar to those reported by \citet{veilleux05} for more luminous active galaxies, as well as to those obtained by \citet{barbosa09} for other nearby Seyfert galaxies and to that obtained by \citet{crenshaw07} for the Seyfert galaxy NGC4151.

\subsubsection{Feeding vs. feedback}

Although the are similarities between the velocity fields observed for  H$_2$ and Br$\gamma$, there are also differences in the detailed kinematics for the two gas phases, as follows. (1) The small but systematic excess blueshift in the H$_2$ emitting gas along PA$=293(-67)^\circ$, suggests there could be an inflow of molecular gas along a nuclear bar, as discussed above. (2) The Br$\gamma~\sigma$ map (Fig.~\ref{gas}) shows the highest values to the S--SE of the nucleus, approximately at the same location where the highest
blueshifts are observed (see Fig.~\ref{res}). This result supports an origin for at least part of  the Br$\gamma$ emission of this region in disturbed gas which is not confined to the galaxy plane but is outflowing from the nucleus along the ionization cone seen in [O{\sc\,iii}]. The H$_2~\sigma$ map, on the other hand,  presents  overall lower $\sigma$ values consistent with a dominant origin in molecular gas orbiting in the plane of the galaxy. (3) The velocity channel maps shown in Figs.~\ref{slices_br} and \ref{slices_h2} also support a larger contribution from  outflowing gas  to the Br$\gamma$  than to the H$_2$ emission, since the blueshifted emission in Br$\gamma$
reaches larger velocity values and is found farther away from the line of nodes than in the H$_2$ emitting gas. 


Previous studies have also found that the H$_2$ traces less disturbed kinematics  than 
the ionized gas in the circumnuclear region of Seyfert galaxies \citep{sb99,rodriguez-ardila04,rodriguez-ardila05,riffel06,riffel08}, leading to the suggestion that the molecular gas, orbiting in the plane, traces the feeding of the active nucleus via inflow of cold gas, while the ionized gas traces its feedback via the outflows. In \citet{riffel08} we could trace streaming motions in the H$_2$ emitting gas towards the nucleus along a nuclear spiral arm. In the present observations, there is also some evidence of inflow along a nuclear bar. The compactness and high inclination of the circumnuclear ring of star-formation does not favor the detection of possible additional streaming motions between the ring and the nucleus. The H$_2$ flux distribution shows that the molecular gas is not concentrated in the star-forming ring, but is distributed over the whole nuclear region, indicating the presence of similar density of molecular gas at the ring and between the nucleus and the ring. This molecular gas could be the material which is feeding and probably will continue to feed the active nucleus for some time in the future.

\subsubsection{AGN-Starburst connection in NGC\,7582}

The total mass of hot H$_2$ in the nuclear region of NGC\,7582 can be estimated as \citep{riffel08}:
\begin{equation}
\left(\frac{M_{\rm H_2}}{\rm
M_\odot}\right)=2.40\times10^{14}\left(\frac{F_{\rm
H_{2}\lambda2.2235}}{\rm
erg\,s^{-1}\,cm^{-2}}\right)\left(\frac{d}{\rm Mpc}\right)^2,
\label{mh2}
\end{equation}
where we assume a vibrational temperature of $T_{vib}=2000$\,K, which
implies a population fraction $f_{\nu=1,J=2}=3.7\times10^{-3}$
\citep{scoville82}, and a transition probability
$A_{S(0)}=2.53\times10^{-7}$\,s$^{-1}$ \citep{turner77}. Integrating the flux of the H$_2\lambda2.2235\,\mu$m emission line over the wole IFU field we obtain 
$F_{\rm H_{2}\lambda 2.2235} \approx 5.5\times 10^{-15}\,{\rm erg\,s^{-1}cm^{-2}}$ and thus  $M_{\rm H_2}\approx620\,{\rm M_\odot}$.

The above value is the mass of hot H$_2$ emitting gas, which is probably only the ``hot skin'' of  the molecular gas available in the nuclear region to feed the active nucleus. The total mass of molecular gas should be much higher, as it should be dominated by cold gas, which has been estimated to be 10$^{5}$ to 10$^{7}$ times that of the hot gas in galaxies for which both hot and cold molecular gas have been observed \citep{dale05}. Thus, there should be plenty of molecular gas, not only to feed the AGN but also to give origin to the recent bursts of star-formation in the ring.

The molecular gas accumulated in the nuclear region was probably the fuel which gave origin to the nuclear ring of star formation, as well as to the present episode of nuclear activity. Although both the star formation and nuclear activity could have been triggered simultaneously, there is also the possibility that the star formation was triggered first, and then the mass loss from stellar winds and supernova explosions in the CNSFRs  have been the source of fuel for the SMBH.
It is thus instructive to estimate the available mass to feed the SMBH under the above assumption. For solar metallicity and an age of 5\,Myr, the total mass loss for the two CNSFRs is in the range $\approx$10$^3-10^5$\,M$_\odot$  \citep{lh95}. Assuming a value of 10$^4$\,M$_\odot$, and that the duration of this process is a few 10$^6$\,yr, one gets a rate of mass loss of $10^{-2}$ M$_\odot$\,yr$^{-1}$. (Note that this value is $\approx$5 times larger than the nuclear accretion rate calculated above, of $\dot{m}\approx2.6\times10^{-3}\,{\rm M_\odot\,yr^{-1}}$, within the uncertainties of the calculations.) If this mass reaches the nucleus along a nuclear bar (as observed in NGC\,7582) or nuclear spiral arms [as observed in NGC\,1097 \citep{fathi06} and NGC\,6951 \citep{sb07}], with velocities of $\approx$20$-$60\,km\,s$^{-1}$, this gas would take  a few $10^{6}$\,yr to reach the nucleus. This timescale alows for the coexistence of the nuclear activity and star-forming regions of 5\,Myr under the hypothesis that the nuclear activity has been triggered  by mass accretion originated from mass loss of young stars formed in the  circumnuclear ring. Alternatively, the triggering may have happened during a previous burst of star-formation in the ring. The presence of such previous starburst is supported by the partial ring of low $\sigma_*$ observed in the stellar kinematics. In addition, as pointed out in Sec.\,\ref{nuclear}, the extranuclear spectrum does not differ significantly from those of the star-forming ring, suggesting that there may be regions of recent star formation even closer to the nucleus, which may be contributing to the SMBH feeding as well.

Finally, we quantify the AGN-Starburst connection in NGC\,7582 by comparing the mass accretion rate to the SMBH with the star formation rate (SFR) in the circumnuclear region. As pointed out in the introduction, the M$-\sigma$ relation implies that the mass of the SMBH in galaxies grows in proportion to the growth of the bulge. If the bulge grows via star formation episodes, the mass accretion rate to the SMBH should be proportional to the SFR. Which should be the ratio between the mass accretion rate and the SFR? The same as the between the mass of the SMBH and the mass of the host galaxy bulge, which was first obtained by \citet{magorrian98}, and today known as the Magorrian relation \citep{fm01}, which has been revised to a ratio of 0.1\% between the mass of the SMBH and that of the bulge.
Assuming that the total SFR can be represented by that of the ring, we get for NGC\,7582 a ratio of 0.26\% between the mass accretion rate and the SFR, a number which is close to that of the Magorrian relation, considering the uncertainties. These uncertainties include, in particular, a possible underestimate of the SFR in the nuclear region, as our observation misses the top border of the star-forming ring, and a possible contribution of a starburst included within the nuclear aperture.

\section{Summary and conclusions}

Two-dimensional near-IR $K-$band spectroscopy from the inner
660$\times$315\,pc$^2$ of the Seyfert galaxy NGC\,7582, obtained with
the Gemini GNIRS IFU at a spatial resolution of $\approx40$\,pc$^2$ and
spectral resolution of $\approx$3\,\AA, was used to map the molecular and ionized gas emission-line flux distributions and kinematics as well as the stellar
kinematics. The region covered by the observations include most of a circumnuclear ring of star-formation in the plane of the galaxy plus a small part of an outflow extending to high galactic latitutes.

Our main conclusions are: 
\begin{itemize}

\item The nucleus contains an unresolved source whose continuum  is well reproduced by a blackbody function with temperature T$\approx$1050\,K, which we atribute to emission by circumnuclear dust heated by the AGN, with an estimated mass of $\approx3\times\,10^{-3}$\,M$_\odot$ located within 25\,pc of the nucleus. The nuclear spectrum shows also a broad component in the Br$\gamma$ emission line with FWHM$\approx$3900\,km\,s$^{-1}$. 

\item The Br$\gamma$ flux distribution is dominated by emission from a circumnuclear ring of star formation, previously seen only in mid-IR observations, with radius of $\approx$\,190\,pc, which contributes with $\approx$80--90\% of the Br$\gamma$ emission, the rest being contributed by the nuclear source. Two large star-forming  regions with age $\approx$\,5\,Myr contribute with 50\% of the Br$\gamma$ emission of the ring, resulting in $\approx$1000 O6 stars per region and a star-forming rate of $\approx$0.25\,M$_\odot$\,yr$^{-1}$.
The total rate of ionizing photons emitted by the ring is $\approx10^{53}$\,s$^{-1}$, the total star-forming rate is $\approx$1\,M$_\odot$\,yr$^{-1}$ and the total mass of ionized gas is $\approx3\,\times\,10^6$\,M$_\odot$. 

\item The H$_2$ flux distribution is more uniformly distributed over the nuclear region than the Br$\gamma$ flux distribution and gives a mass of hot molecular gas of $\approx$\,620\,M$_\odot$. From previous studies, it can be concluded that the total mass of H$_2$ gas (which is dominated by cold H$_2$) can be 10$^5$--10$^7$ times larger.

\item The stellar velocity field shows a distorted rotation pattern whose centre appears to be displaced up to 50\,pc from  the nucleus (identified as the peak of the continuum emission). The distortion could be  associated to a nuclear bar and a nuclear spiral previously seen in mid-infrared images. The velocity dispersion of the bulge is $\sigma_*\approx170\,{\rm km\,s^{-1}}$, implying in a mass of $M_{\rm BH}\approx7\times10^7\,{\rm M_\odot}$ for the SMBH, in
good agreement with previous determinations. Immersed in this bulge, there is a partial ring of lower $\sigma_*$ ($\approx50\,{\rm km\,s^{-1}}$), which is close to the star-forming ring but displaced $\approx$\,50\,pc inwards. As the velocity dispersion is similar to that of the molecular gas, this ring can be interpreted as due to stars recently formed from cold gas, although in a previous burst (age\,$\ge$\,10$^7$\,yr), still keeping the gas kinematics as they did not have time yet to ``thermalize'' with the stars of the bulge.

\item The  radial velocity field of the ionized gas is similar to that of the stellar component to the N--NW, while to the S--SW there is an additional blueshifted component ($v\approx-100$\,km\,s$^{-1}$), which we attribute to outflows along the ionization cone (previously observed in a narrow-band [O{\sc\,iii}] image) which is partially covered by our observations. Velocity channel maps along the Br$\gamma$ emission-line profile show even higher blueshifts, reaching $v\approx-300$\,km\,s$^{-1}$ in the outflow, while the maximum redshift observed is $v\approx200$\,km\,s$^{-1}$. The velocity dispersion is enhanced in the region of the outflow.

\item
The mass outflow rate in the ionized gas is estimated to be 
$\dot{M}_{\rm HII}\approx0.05\,{\rm M_\odot\,yr^{-1}}$, which is an order of magnitude larger than the accretion rate to feed the AGN, a ratio which is similar to those found in other AGN. This large ratio indicates that the outflowing gas does not originate in the AGN, but is instead the circumnuclear gas from the host galaxy being pushed away by a nuclear outflow.

\item The kinematics of the hot molecular gas, traced by the H$_2$ emission, shows smaller blueshifts along the outflow as well as lower velocity dispersions, which suggests that most of the molecular gas is in the galactic plane. An excess blueshift along  PA$\approx-70^\circ$, where a nuclear bar has been observed, can be interpreted as an inflow towards the nucleus. We thus conclude that the H$_2$ kinematics traces the feeding of the AGN, while the ionized  gas kinematics traces its feedback via the outflows.

\item
The estimated mass loss rate from the evolving stars in the circumnuclear ring is $\approx10^{-2}$\,M$_\odot$\,yr$^{-1}$, which is  $\approx$\,5 times the nuclear accretion rate. As the estimated time for this mass to reach the nucleus is a few 10$^6$\,yr, this material could be the fuel which just triggered the nuclear activity.
Nevertheless, as there is also a previous burst of star formation almost co-spatial with the ring, the fuel may have been available previously (e.g. $\approx10^7$\,yr). In addition, previous data suggest that there may be also recent star formation inside the ring, closer to the nucleus.

\item
We conclude that the AGN--Starburst connection in NGC\,7582 may have occurred in two ways: (1) there is a large molecular gas reservoir in the nuclear region which gave origin to both the star-formation in the circumnuclear ring and nuclear activity; (2) the molecular gas has given origin first to circumnuclear star formation and the mass loss from the evolving stars have then triggered the nuclear activity. 

\item
The ratio between the mass accretion rate and the SFR in the circumnuclear region, of 0.26\% is close to the value implied by the Magorrian relation between the mass of the SMBH and the mass of the bulges of the host galaxies. This result shows that a growth of the SMBH proportional to the growth of the bulge can proceed via circumnuclear star formation around AGN.

\end{itemize}

\section*{Acknowledgments}
We thank the anonymous referee for valuable suggestions which helped to improve the paper, as well as Dr. Michele Cappellari for help with the pPXF routine. Based on observations obtained at the Gemini Observatory, which is operated by the Association of Universities for Research in Astronomy, Inc., under a cooperative agreement with the NSF on behalf of the Gemini partnership: the National Science Foundation (United States), the Science and Technology Facilities Council (United Kingdom), the National Research Council (Canada), CONICYT (Chile), the Australian Research Council (Australia), Minist\'erio da Ci\^encia e Tecnologia (Brazil) and SECYT (Argentina).  This work has been partially supported by the Brazilian
institution CNPq.

\label{lastpage}

\end{document}